\documentclass[fedsstyle, oneside, onecolumn, Numbered]{jnl}

\usepackage{graphicx}
\usepackage{amsmath,amssymb,amsfonts}
\usepackage[title]{appendix}
\usepackage{manyfoot}
\usepackage{booktabs}
\usepackage{bibunits}
\usepackage{titletoc}
\usepackage{lmodern}
\usepackage{subfigure}

\defaultbibliographystyle{fedsstyle}

\hypersetup{urlcolor=black}

\newenvironment{aligned*}{
\begin{equation}
\begin{aligned}}{%
\end{aligned}%
\end{equation}}

\title{
Confinement of hybridized light modes in nonlinear-optical superpotentials
}

\author[1]{\fnm{Federico} \sur{Presutti}}\email{fp267@cornell.edu}
\presentaddress{Research Laboratory of Electronics, Massachusetts Institute of Technology, Cambridge, MA, USA}
\author[1,2]{\fnm{Ryotatsu} \sur{Yanagimoto}}\email{ry338@cornell.edu}

\affil[1]{%
 \orgdiv{School of Applied and Engineering Physics},
 \orgname{Cornell University},
 \orgaddress{%
   \state{NY}%
   , \country{USA}}%
}
\affil[2]{%
 \orgdiv{NTT Physics and Informatics Laboratories},
 \orgname{NTT Research, Inc.},
 \orgaddress{%
   \state{CA}%
   , \country{USA}}%
}

\begin{document}
\begin{bibunit}

\abstract{
In this work, we present a spatial and temporal light trapping mechanism induced by coherent frequency conversion.
We consider a wave-mixing process, where a pumped nonlinearity mediates a frequency-domain beamsplitter interaction between two signal modes.
When these wave modes are counterpropagating, temporally trapped eigenmodes emerge at the zero of the nonlinearity profile.
We establish that such spatio-temporal trapping has a direct connection to supersymmetric quantum mechanics, which ensures the existence and robustness of the trapped modes and allows us to derive analytic solutions of the bound eigenmodes.
Backwards sum-frequency conversion may thus enable a unique class of active photonic memory devices, whose design parameters are compatible with the state of the art in integrated nonlinear photonics.
We propose proof-of-principle experiments to study this nonlinear-optical trapping effect on thin-film lithium niobate.
In addition, as a case study, we show how the temporal trap can be used alongside parametric gain.
Our work unveils potential applications ranging from classical integrated photonics to multimode quantum nonlinear photonics.
}

\phantomsection
\addcontentsline{toc}{section}{Abstract}

\maketitle

\phantomsection
\addcontentsline{toc}{section}{Introduction}

Wave localization is a ubiquitous yet profound phenomenon, playing a crucial role across diverse areas of physics.
While the most basic model of localization involves a wavepacket contained in a medium where the phase velocity is slower than the surroundings, the ``particle in a potential'' model,
more intricate mechanisms of wave confinement have been discovered, including Anderson localization and Aharanov--Bohm caging \cite{Anderson1958, AharonovBohm}, bound states in the continuum~\cite{HsuSoljacic, vonNeumann1993}, and topological edge states~\cite{ThoulessNijs, Hatsugai}.
Arguably, photonics has been one of the most successful platforms not only for demonstrating such wave-localization physics~\cite{Segev2013, BandresKhajavikhan, KodigalaBoubacar, DuttFan, MukherjeeGoldman},
but also in leveraging it to realize indispensable technologies such as waveguides, resonators~\cite{Vahala}, and optical fiber technologies~\cite{ChenTextbook}.
Indeed, there are wave-trapping mechanisms that are unique to photonics, such as gain guiding via imaginary index contrast \cite{SalinSquier, Siegman}; photonic crystal cavities and waveguides, where the photonic bandgap confines light to defects \cite{Russell, Yablonovitch}; and light with high orbital angular momentum can be trapped by a centrifugal barrier \cite{MaRamachandran}.
Beyond waveguiding with fabricated material structures,
nonlinear interactions enrich the wave-trapping picture, mediating the localization of one field by another -- or by itself~\cite{IslamGordon, AgrawalTextbook, WangKudlinski, LocalizedWavesBook} -- enabling mode locking and soliton formation~\cite{GreluAkhmediev}.

\begin{figure}[!hbtp]
\centering
\subfigure{\label{fig:applications:A}}
\subfigure{\label{fig:applications:B}}
\subfigure{\label{fig:applications:C}}
\subfigure{\label{fig:applications:D}}
\includegraphics[width=0.86\normaltextwidth]{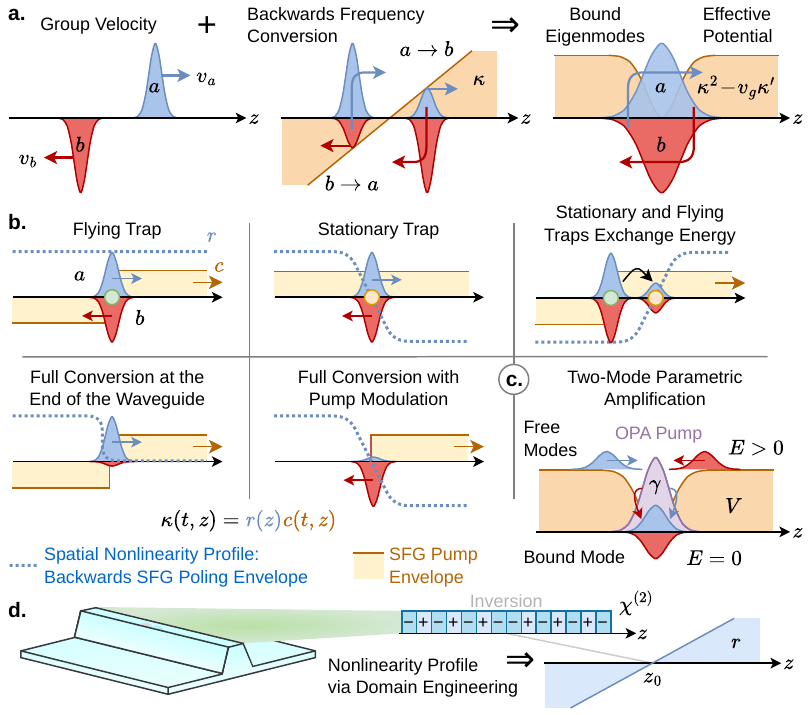}
\caption{
 \textbf{Coherent photonic traps via sum-frequency generation.}
 \textbf{a.} Group velocity and frequency conversion induce bound eigenmodes.
 Two modes $a,b$ at different wavelengths walk off with group velocities $v_a, v_b$ in opposite directions.
 The modes are coupled via a frequency conversion process with a spatially varying profile $\kappa$.
 Bound eigenmodes form around the position where $\kappa$ crosses 0.
 The trapping is a result of the balance between the walk off and the conversion reversing the group velocity back towards the center point.
 \textbf{b.} Flying and stationary traps.
 Flying traps are characterized by a pump that co-propagates with the two trapped modes $a, b$.
 The pump's temporal profile $c$ must have a zero with phase inversion.
 A stationary mode is fixed in space, with a constant pump in time, but with an appropriate spatially varying nonlinearity profile $r$, e.g.\ via the $\chi^{(2)}$ domain poling, as illustrated in \textbf{d}.
 A stationary trap can exchange energy with a flying trap as it traverses it.
 At the end of a nonlinear waveguide where the mode coupling vanishes, flying mode $b$ is completely converted into mode $a$.
 Similarly, rapidly switching off $c$ causes a complete conversion of stationary $a$ into $b$.
 \textbf{c.} Two-mode optical parametric amplification (OPA) on the eigenmodes of the trap.
 The two-mode OPA process shares an eigenbasis with the trap.
 The ground state mode has zero energy $E$ (unchanged phase velocity), hence remains phase matched, while free and higher-order modes have an additional energy dependent on the depth of the trap.
}
\label{fig:applications}
\end{figure}

In this work, we introduce a trapping mechanism of waves based on the interplay between group-velocity mismatch and mode-coupling interactions.
We are specifically interested in the quasi-phase matched backward three-wave mixing process,
and we unveil the connection of this trap to the energy potentials of supersymmetric quantum mechanics (SUSY QM) \cite{CooperKhareSukhatme}.
We consider two counter-propagating optical modes interacting via sum- and difference-frequency generation.
Intuitively, the frequency conversion process that reverses the propagation direction of light acts analogously to a pair of ``mirrors'' reflecting these modes -- which would otherwise walk away -- back towards each other and mimicking a Fabry--P\'erot resonator.
Alternatively, in the ``particle inside a potential'' picture, the conversion acts as a restoring force, counteracting the ``momentum'' due to group velocity.
This active trapping mechanism follows similar physics to (passive) distributed Bragg reflectors \cite{Longhi}.
Consequently, when the spatial distribution of the coherent coupling traverses zero and reverses phase, a cavity or resonator is formed that supports bound or quasi-bound modes, via the hybridization of the counter-propagating wave modes (see Fig.~\ref{fig:applications:A}).
The unique structure of the Hamiltonian not only allows us to derive analytic expressions for the trapped states but also ensures that the energy of the ground state is pinned to zero, regardless of the system parameters, suggesting a unique trapping mechanism that is robust against noise and imperfections.
Unlike the other demonstrations of SUSY in photonics that require precise fabrication of perfectly matched pair-potential structures
\cite{GarciaMecaLlorente,ChumakovWolf,MiriChristodoulides,ElGanainyChristodoulides,HokmabadiKhajavikhan,QiaoFeng,HeinrichChristodoulides,YimLiang,MiriChristodoulides2,MachoGarciaMeca},
the emergent SUSY behavior in our system is tied to the topological structure of the mode-to-mode interaction and thus is uniquely robust.

After deriving the theory, we verify through numerical simulations the ability of the system to realize a cavity or a resonator on demand, controlled by modulation of the pump light.
We then perform a feasibility study of the proposed scheme on thin-film lithium niobate nonlinear nanophotonics, showing that the requirements for a high-fidelity
implementation of the trapping mechanism is well within the state of the art.
We discuss potential applications and implications of such implementations.
Namely,
we distinguish between two types of coherent traps that can be realized along the length of a nonlinear waveguide: stationary and flying.
The former is based on a spatially varying nonlinearity -- i.e.\ the quasi-phase-matching (QPM) profile -- while the latter is based on temporal pump modulation and propagates at the speed of the pump wave.
The flying cavities can traverse stationary ones, and mediate spatial unitaries between them or couple light in from outside the device.
This may enable a unique class of sought-after photonic memory devices, or optical buffers
\cite{YanikFan, YanikFan2, YanikSuhFan, YanikFan3, ReedSoljacicJoannopoulos}, significantly broadening the scope of ``how light can trap light.''
In principle, the devices would require no additional geometries other than the waveguide itself, in addition to the material poling for QPM and an external pump modulation mechanism.
Finally, in the quantum-optical context,
a natural application of our result is to enforce single- or few-mode behavior on naturally multimode dynamics.
As a concrete example, we show how single-mode squeezing
can be realized, providing a new avenue of research to approach widely recognized challenges in quantum science and technologies~\cite{HoudeQuesada, UrenRaymer, AnsariSilberhorn, QuesadaSipe}.

\section*{Results}\label{sec:resultsAFCOPA}
\addcontentsline{toc}{section}{\nameref{sec:resultsAFCOPA}}

\subsection*{Three-wave Mixing and the Emergent Energy Landscape}

For detailed background and theory we direct the reader to Appendix~\ref{sec:app:theoreticalbackground}, as here we limit ourselves to summarizing the results.
Although the general phenomenon we describe may apply to any coupled-wave system, henceforth, we consider an experimentally relevant case of backwards sum-frequency generation (SFG) in a photonic waveguide.

We consider two optical modes in a nonlinear waveguide, denoted by $a(t,z)$ and $b(t,z)$, where $t, z$ are the time and spatial coordinates.
Importantly, we assume that the field envelopes $a$ and $b$ propagate in the $+z$ and $-z$ directions, with group velocities $v_a$ and $v_b$, respectively.
In the absence of nonlinear interactions, the dynamics are governed by equations
\begin{aligned*}
\frac{da}{dt} &= v_a \frac{da}{dz} , \\
\frac{db}{dt} &= -v_b \frac{db}{dz} .
\end{aligned*}

We now introduce a pump field with amplitude $c(t,z)$ propagating in the $+z$ direction, which couples the modes $a$ and $b$ through a three-wave mixing process, such that the three frequencies satisfy $\omega_a = \omega_b + \omega_c$
(for clarity we henceforth assume the case $\omega_a > \omega_b$).
We assume the undepleted pump approximation and that this process is quasi-phase matched.
In this process, a photon in mode $a$ interacts with a pump photon and converts into mode $b$, and vice versa, thereby propagating in the opposite direction.
The interplay between frequency conversion and group-velocity mismatch is illustrated in Fig.~\ref{fig:applications:A}.
By including this mode coupling, the equations of motion become
\begin{aligned*}
\frac{da}{dt} &= -\kappa(z) b + v_a \frac{da}{dz} , \\
\frac{db}{dt} &= \kappa(z) a - v_b \frac{db}{dz} .
\label{eq:coupledmodesMain}
\end{aligned*}
The coupling strength is given by $\kappa(t, z) = r(z) c(t, z)$, where $r(z)$ is proportional to the nonlinear coupling coefficient of the waveguide,
which, in addition to the material nonlinearity,z is defined by the $\chi^{(2)}$ domain poling profile.
A profile of $r(z)$ can be fashioned by domain engineering ferroelectric nonlinear $\chi^{(2)}$ materials (Fig.~\ref{fig:applications:D}).

Based on the connection to supersymmetric quantum mechanics highlighted in Appendix~\ref{sec:app:theoreticalbackground}, we may call $\kappa$ the superpotential.
The superpotential induces an effective potential energy landscape, dictating a basis of bound eigenmodes, which are hybridized modes of $a$ and $b$.
We may define two superpositions:
\begin{aligned*}
\alpha_- &= \sqrt{v_b} a - \sqrt{v_a} b , \\
\alpha_+ &= i \left( \sqrt{v_b} a + \sqrt{v_a} b \right) .
\end{aligned*}
The weights indicate that more time is spent in the slower mode (for $v_a, v_b$ in the frame of reference where $\kappa(z)$ is stationary).
To symmetrize the equations, we define $v_g = \sqrt{v_a v_b}$, the characteristic group velocity.
By taking the linear combinations of Eq.~\ref{eq:coupledmodesMain}, we can rewrite the equations as
\begin{aligned*}
\frac{d}{dt} \alpha_+ = A \alpha_-, & \hspace{0.6em}
\frac{d}{dt} \alpha_- = A^\dagger \alpha_+, \\
\text{with } A &= \kappa(\zeta) + \frac{d}{d\zeta} .
\end{aligned*}
$A, A^{\dagger}$ are generalized ladder operators associated with the superpotential under normalized coordinate $\zeta = z / v_g$, reducing to the familiar harmonic-oscillator ladder operators when $\kappa \propto \zeta$.
Our equations are thus formally equivalent to SUSY QM.
Applying the two relations in succession decouples the hybridized modes, which we can formulate as a pair of Schr\"odinger-type eigenproblems, with eigenvalues $\lambda^2$ and eigenmodes $\alpha_\pm^{(\lambda)}$:
\begin{aligned*}
\frac{d^2}{dt^2} \alpha_\pm
&= \left\{\begin{matrix}
A A^\dagger \\
A^\dagger A
\end{matrix}\right.
\alpha_\pm 
= H_\pm \alpha_\pm \\
\lambda^2 \alpha_\pm^{(\lambda)}
&= \left( -v_g^2 \frac{d^2}{dz^2} \pm V_\pm(z) \right) \alpha_\pm^{(\lambda)}
, \\
\text{with } V_\pm &=
\kappa^2 \pm v_g \frac{d\kappa}{dz}
.
\end{aligned*}
We use that the operators in the conditional brace (\{) are aligned with the $\pm$.
In the language of SUSY QM, $H_\pm$ are the superpartner Hamiltonians, whose spectra coincide for all $\lambda \neq 0$.
The field for each hybridized mode is thus subject to the potentials $V_\pm$, and has an energy (phase velocity) of $\lambda^2$.
Notably, an effective mass (a momentum-like term in the Hamiltonian) emerges even though the initial model contained only group velocities.
Indeed, the trapping arises because the conversion acts similarly to a restoring force, balancing the ``momentum'' of the group velocity.
In such a trap eigenstate, the photons are in a superposition of two different colors (wavelengths) traveling backward and forward, with no net propagation:
these massless photons form a fictitious massive particle due to their coupling, allowing stationary, localized states.

In order for $\kappa$ to be a valid superpotential, it must be real-valued (up to some global phase, for Hermiticity) and have opposite sign as $z \rightarrow \pm \infty$, hence must have a point where $\kappa(z_0) = 0$ corresponding to a $\pi$ phase (sign) flip.
This condition guarantees that the ground state of the potential is real-valued and has zero energy, $\lambda = 0$, which means that the phase velocity of the underlying $a, b$ waves are unchanged.
This is referred to as unbroken SUSY.
Because this is a topological condition on the zero and asymptotics of $\kappa$ rather than on its shape or amplitude, the existence of a ground state with an eigenvalue of exactly zero is robust against pump-power fluctuations and profile imperfections.

As an example, let us consider the well-known tanh superpotential \cite{CooperKhareSukhatme}, which yields the sech$^2$ (P\"oschl--Teller) potential.
Notably the number of bound states is analytically determined by the depth.
Specifically,
\begin{equation}
\kappa(z) = K_m \tanh \frac{z-z_0}{l_0}
\end{equation}
for some length scale $l_0$ on the order of $v_g / \kappa$, has $m$ bound states when
$K_m = m |v_g| / l_0$,
for integer $m$.
Hence, trapping allows us to define a robust subspace of a variable number of modes.

In the following sections we present potential techniques to harness the trapping mechanism, by control of the nonlinearity profile in space or time (Fig.~\ref{fig:applications:B}).
In addition, we will discuss how an effective potential may be combined with parametric amplification to add gain to the trap (Fig.~\ref{fig:applications:C}); we show that localized, single- or few-mode squeezing could be achieved within the potential.
Finally, we explore the possibility of an experimental realization and present a blueprint for a proof-of-concept photonic device.

\subsection*{Stationary and Flying Traps}

\begin{figure}[!htbp]
\centering
\subfigure{\label{fig:traphopping:A}}
\subfigure{\label{fig:traphopping:B}}
\includegraphics[width=1.0\normaltextwidth]{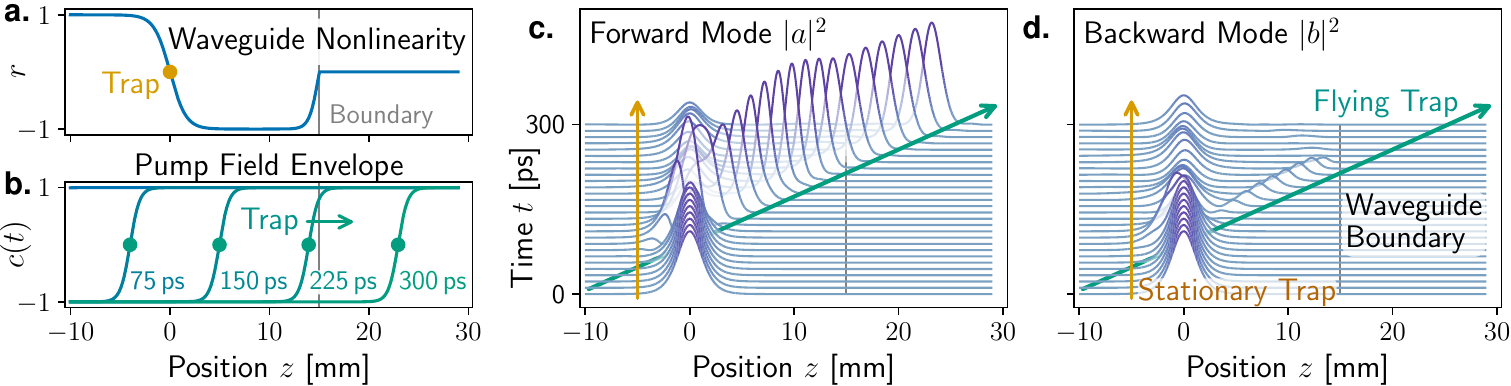}
\caption{
 \textbf{Coupling into and out of stationary traps via flying traps.}
 \textbf{a.} Waveguide nonlinearity profile.
 A 30\,mm waveguide with a tanh profile nonlinearity was simulated.
 $z=0$ represents the center of the waveguide.
 \textbf{b.} Pump field envelope profile as a function of time.
 A tanh pump profile was considered.
 The modulation propagates with the group velocity of the pump.
 \textbf{c.} The forward mode intensity profile in the waveguide over time.
 Starting in the stationary ground eigenstate, energy is coupled from the waveguide trap into the flying trap as the pump profile's modulation passes through.
 \textbf{d.} The backward mode intensity in the waveguide over time.
 The backward mode accounts for a lesser part of the superposition in the flying trap, due to the greater difference in group velocity with respect to the pump.
 Once the spatial nonlinearity vanishes, the backward mode in the flying trap is converted into the forward mode.
 }
\label{fig:traphopping}
\end{figure}

As touched on in the previous section,
trapping can occur if the group velocity of the nonlinearity profile $\kappa$ lies between those of the modes $a,b$.
In nonlinear optics, the coupling profile is composed of both the material nonlinearity $r$ and the pump field envelope $c$.
Either may be used to induce a trap,
and are illustrated in Fig.~\ref{fig:applications:B}.
A temporally modulated pump over a uniform nonlinearity can trap co-propagating modes;
the time-dependent nonlinearity propagates at the pump group velocity, in which case we call it a ``flying'' trap.
A zero and a corresponding phase inversion may be designed into the spatial nonlinearity profile (the $\chi^{(2)}$ domain poling); under a continuous-wave (CW) pump: this is a stationary trap.
The group velocity condition for $a,b$ is more strict, as the wave mixing must be backwards SFG: $v_b < 0 < v_a$,.

The stationary trap is the more interesting case as it realizes an active optical cavity, holding a pulse of light along an otherwise spatially uniform waveguide, which persists as long as it is pumped.
A flying trap only exists while the modulation traverses the waveguide, hence is not necessarily useful \textit{per se} as a cavity.
Nonetheless, flying traps may serve to couple light into and out of the stationary traps and the nonlinear waveguide without resorting to any other external coupling device,
as illustrated in Fig.~\ref{fig:applications:B}.
More generally, they may be used to mediate spatial unitaries to move light between stationary traps along the same waveguide.
If a flying trap reaches the end of the nonlinear waveguide, full conversion will occur into the forward mode (in the backward SFG case);
similarly, rapidly switching off the pump to a stationary trap will cause full conversion into the backwards mode.

Fig.~\ref{fig:traphopping} presents a simulation demonstrating a state bound in a stationary trap partially coupling into a flying trap, which then exits the nonlinear waveguide, releasing its state into the forward mode.
The velocity parameters used in the simulation were $v_a = -v_b = c_0 / 2$ and $v_c = c_0 / 2.5$.
A $\tanh(z / l_0)$ profile was used as the nonlinearity $r$ and modulated pump profile $c$, with length scales of $l_0 =$ 1600 and 950\,{\textmu}m respectively, within a 30\,mm nonlinear waveguide.

\subsection*{Trapping to Enforce Single-mode Parametric Amplification}

\begin{figure}[!htbp]
\centering
 \subfigure{\includegraphics[width=0.33\normaltextwidth]{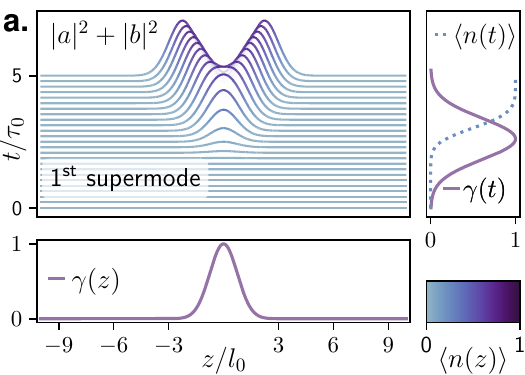}\label{fig:squeezetrap:A}}
 \subfigure{\includegraphics[width=0.33\normaltextwidth]{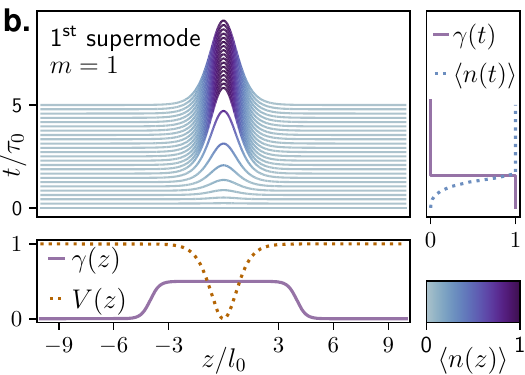}\label{fig:squeezetrap:B}}
 \subfigure{\includegraphics[width=0.33\normaltextwidth]{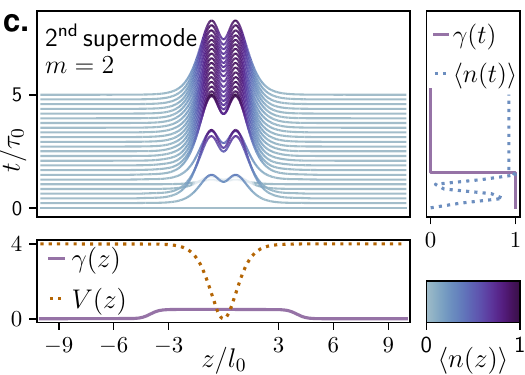}\label{fig:squeezetrap:C}}
\caption{
 \textbf{Squeezed vacuum with a single mode trap.}
 We compare the photon number profile $\langle n \rangle$ of a given supermode throughout the vacuum squeezing process, with and without trapping.
 We used normalized coordinates and variables, with timescale $\tau_0 = l_0 / v_g$.
 \textbf{a.} Conventional two-mode squeezing, with a Gaussian pump profile, including in the propagation direction (i.e. apodization).
 Group velocity mismatch is a necessary condition to achieve a minimal Schmidt number \cite{UrenRaymer}, hence the supermode separates as its two components walk off.
 \textbf{b.} Squeezing with a single-mode sech$^2$ potential.
 The mode's profile remains unchanged throughout the squeezing process as well as after the gain is turned off ($\gamma \rightarrow 0$).
 \textbf{c.} Squeezing with a deeper, two-mode potential.
 The evolution of the second trap eigenmode is shown.
 The profile also remains fixed.
 However, as this mode sits at a higher energy in the potential, it is not phase matched, hence the gain oscillates between amplification and de-amplification.
}
\label{fig:squeezetrap}
\end{figure}

Temporal bound modes have been considered in combination with operations that are not linear-optical, such as optical parametric amplification (OPA) or squeezing in order to mitigate unwanted multimode behavior \cite{YanagimotoMabuchi}; however, our coherent trapping mechanism represents an entirely new approach.
In particular, since the wave coupling and the two-mode parametric amplification Hamiltonians may share an eigenbasis (depending on the OPA pump profile) the two processes may be easily combined, so that the bound modes are independently amplified.
(Fig~\ref{fig:applications:C}. See Appendix~\ref{sec:app:opahamiltonian} for details.)

Concretely, the simultaneous OPA would contribute the terms $i \gamma(t, z) b^*$ and $i \gamma(t, z) a^*$ to the equations of motion given in Eq.~\ref{eq:coupledmodesMain}, where $\gamma$ is the parametric amplification drive.
Hence, pumping a uniform $\gamma$ over the trap would retain the eigenmodes (``supermodes'') of the unpumped trap.
Furthermore, OPA amplifies the odd and even superpositions of $a, b$:
because the trap already imposes the bound modes to be in these superpositions, it is possible to directly perform single-mode squeezing without requiring a beamsplitter operation to unentangle two states.

Simulations of trapped-mode vacuum squeezing are presented in Fig.~\ref{fig:squeezetrap} in normalized units, for simplicity.
We consider the tanh superpotential to maintain a fixed number of squeezing supermodes \cite{WasilewskiRadzewicz}, which do not change throughout propagation.
Any modes of higher order $m$ rotate harmonically with additional phase $\propto m$.
As a comparison, panel \subref{fig:squeezetrap:A} simulates the conventional method of two-mode squeezing, using a Gaussian parametric amplification profile $\gamma$ and apodization \cite{UrenRaymer}.
The supermode spreads out with propagation distance, due to the requisite group velocity mismatch.
In contrast, when the superpotential is present, such as in panels \subref{fig:squeezetrap:B}--\subref{fig:squeezetrap:C}, the trapped supermodes remain fixed.
Panel \subref{fig:squeezetrap:B} shows the simulation in the case of a single-mode trap, while \subref{fig:squeezetrap:C} shows the case of a deeper, two-mode trap, and the propagation of the higher-order mode.
We employ a flat-top parametric amplification profile $\gamma$ that is as wide as the potential, to match the squeezing supermodes and the potential's eigenmodes.
We assume that the parametric gain profile $\gamma$ is stationary with respect to $\kappa$.

Realizing a potential that has a single or few bound modes does not necessarily prevent the squeezing of unbound supermodes.
However, their energy will be greater, and will thus not be phase-matched, or their frequency will be detuned from the bound modes.

Beyond squeezing, classical-regime OPA is a possibility, with the distinction that the trap can be arbitrarily shaped and Q-switched.
We note two other forward--backward wave systems with gain: distributed feedback lasers \cite{Longhi} and mirror-less optical parametric oscillators \cite{Harris, CanaliasPasiskevicius}.
The former is passive, using distributed Bragg reflectors as leaky traps to lase; and the latter has no trapping mechanism, but oscillates due to the parametric gain between counter-propagating signal and idler.

\subsection*{Experimental Feasibility of Trapping via Quasi-phase Matching of Backwards Wave Mixing}

\begin{figure}[!htbp]
\centering
 \subfigure{\includegraphics[width=0.305\normaltextwidth]{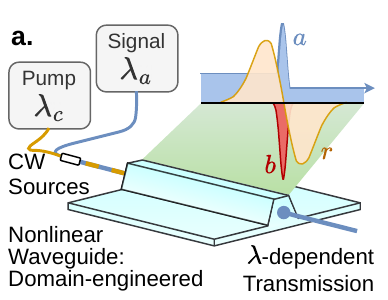}\label{fig:resonator:A}}
 \subfigure{\includegraphics[width=0.695\normaltextwidth]{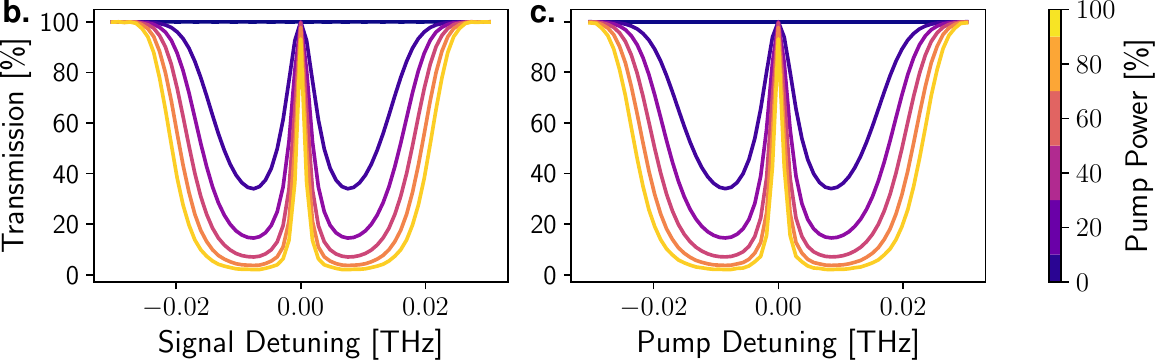}\label{fig:resonator:B}}
 \subfigure{\label{fig:resonator:C}}
\caption{
 \textbf{Hitting a nonlinear-optical trap resonance.}
 \textbf{a.} Proposed experimental configuration.
 A nonlinear waveguide is domain-engineered to form a localized nonlinearity profile, such as the first Hermite--Gauss mode.
 A pump and a signal continuous-wave (CW) laser, with wavelengths $\lambda_c, \lambda_a$, at least one of which is tunable, are coupled into the waveguide.
 By sweeping the wavelength, the transmission curve will reveal the quasi-bound mode resonance.
 \textbf{b.} Simulated signal transmission as a function of its detuning.
 The transmission through the resonator was simulated for varying pump power and signal detuning.
 The transmission peaks at the resonance, and drops with detuning and greater pump power.
 When the detuning is high enough that SFG is no longer quasi-phase matched, the barrier becomes invisible and the transmission returns to unity.
 \textbf{c.} Simulated signal transmission as a function of pump detuning.
 The transmission was simulated for a varying pump and constant signal.
 A similar transmission spectrum is obtained.
}
\label{fig:resonator}
\end{figure}

We now discuss possible approaches to an experimental prototype and the design considerations for the fabrication of a device for coherent trapping.
For clarity, we first connect experimental parameters to the theory models discussed above.
The coupling strength to walk-off ratio in a $\chi^{(2)}$ nonlinear platform, which determines the trap length scale, is given as:
\begin{equation}
\kappa / v_g = l_0^{-1} = \sqrt{\eta P} ,
\end{equation}
where $P$ is the pump power, and $\eta$ is the conversion efficiency, typically reported in units of \%\,/\,W\,cm$^2$.
Recent innovations have enabled thin-film lithium niobate (TFLN) waveguides to reach a range of 5,000--10,000 \%\,/\,W\,cm$^2$ \cite{FrankenLoncar, ChenFan} for second harmonic generation (SHG).
So far, up to 10,000 \%\,/\,W\,cm$^2$ has been reported for SFG \cite{WuXia}.
Note that, all things being equal, fundamentally SFG is $\times4$ more efficient than SHG.
Therefore, at current state-of-the-art efficiencies, it is conceivable that a continuous 100\,mW pump laser could achieve a trap length scale of down to $l_0 = v_g / \kappa \approx$ 1500\,{\textmu}m -- a single trap could occupy a few mm section of a waveguide.
The nonlinearity profile $r(z)$ is realized by designing a quasi-phase matching grating using either domain deletion or duty cycle modulation to locally reduce the effective nonlinearity.
Switching the directions of the domains in the grating is equivalent to flipping the sign of $r$.

The quasi-phase matching (QPM) grating for sum-frequency generation has the following momentum-conservation requirement:
\begin{equation}
 \Delta k_\text{QPM} = k_c + k_b - k_a .
\end{equation}
Here $k$ are the wavenumbers in the waveguide, and $k_\text{QPM}$ is the wavenumber of the domain poling grating to phase match the wave mixing process.
In forwards SFG, the QPM must only compensate for material dispersion, as the vacuum wavenumbers cancel for the energy-conserving process.
However, for backwards SFG, the backwards wave has a negative wavenumber, $k_b < 0$.
This term dominates, and
we therefore have that $\Delta k_\text{QPM} \approx k_b$; the poling period has to be on the order of the wavelength of mode $b$.
While the fabrication may be more demanding, backwards conversion via sub-micron poling has been demonstrated by various research groups \cite{SabattiGrange, YangTang, KoyazBres}.

In an experimental realization, at the high level the design consists in the choice of wavelengths and of the material platform.
Several parameters may need to be considered and optimized simultaneously: 1) the group velocities (GV) and dispersion -- unless designing a purely stationary trap, the pump and signal GV condition must be satisfied, and the GV dispersion minimized; 2) the required domain poling -- backwards SFG typically demands sub-wavelength domains, and shorter period gratings may be less reliably fabricated; 3) the availability of light sources at the given wavelengths; and 4) the material transparency and waveguide cutoff or loss at a given wavelength, especially at longer wavelengths.
The group velocities and the domain poling period presents itself as likely the most important trade-offs, as we discuss below.

We now consider a few wavelength combinations on TFLN.
One approach, supposing that a low $\Delta k_\text{QPM}$ is the main consideration, is to minimize the difference between $k_c$ and $k_a$, and choose a long $b$ wavelength.
For example, a 1550\,nm pump could support signal modes at 980\,nm and 2.6\,{\textmu}m.
With a 2.6\,{\textmu}m backwards mode, the quasi-phase matching period becomes micron-sized.
This wavelength is within the TFLN transparency window, and as the backwards mode, it need not be seeded nor detected.
In a normal dispersion material, in the absence of dispersion engineering, the shorter wavelength $a$ will have a slower group velocity than the pump, forbidding flying traps.
If these were desired at this wavelength combination, birefringence or modal phase matching could be employed.
Specifically, the material must have sufficiently strong birefringence and the susceptibility tensor allow wave mixing between different polarizations, e.g.\ Type II SFG in lithium niobate with the pump polarized along the slow axis.
Alternatively, SFG can be designed for phase-matching with a pump in a higher order spatial mode.
These schemes would come at the cost of reduced nonlinearity.

Another possible configuration is that a 1550\,nm pump could trap another 1550\,nm and a 775\,nm mode.
In Type 0 SFG, the 775\,nm must be forwards, but in Type II, a slow-axis-polarized pump is compatible with a slow-axis backwards 775 and fast-axis forward 1550, thus having all telecom wavelengths in the forward direction, and satisfying group velocity condition.
The poling lengths would, however, be sub-micron.
Additionally, a slow-axis 515\,nm pump could be used for OPA.

Finally, we note that a proof of concept of the nonlinear-optical potentials can be realized without full trapping.
The superpotential can instead be fashioned into a resonator by having $\kappa \rightarrow 0$ sufficiently close to the center of the trap, thus supporting quasi-bound modes.
Scattering theory predicts that the transmission is maximized through a positive potential barrier when the incident light is at resonance.
Experimentally, the pump and the signal can be coupled into the nonlinear waveguide, where one of the sources is wavelength-tunable (Fig.~\ref{fig:resonator:A}).
A sweep of the wavelength would reveal a transmission peak when the signal wavelength corresponds to the forward SFG wavelength.

This was simulated and the results are shown in Figs.~\ref{fig:resonator:B}~and~\ref{fig:resonator:C}, for both signal and pump detuning, and as a function of pump power (nonlinearity).
The spatial superpotential was shaped as the 1st Hermite--Gauss function, and the parameters assumed were $v_g = c_0/2$, $\eta = 10^4$\,\%\,/\,W\,cm$^2$, and $P=100$\,mW (at 100\,\% power), corresponding to a length scale $l_0 = 3000$\,{\textmu}m.
Notably, the resonance creates a transmission peak at zero detuning, but at greater detuning, where the SFG process is no longer phase-matched, the transmission returns to unity, as the barrier has no effect on the signal wavelength.
The width of the peak is inversely proportional to $l_0$.
The transmission is also pump-dependent, as this changes the height of the effective energy barrier and the resonances therein,
hence, positive potentials are comparable to active and tunable spectral filters.

\section*{Discussion}

In this work, we identify a unique mechanism of wave trapping that occurs when two wave modes with a relative group-velocity mismatch interact via position-dependent coupling.
Interestingly, this trapping mechanism can be mapped to the theoretical formulation of SUSY QM, which we can leverage to concisely explain unique features of the trap, e.g.,\ the robustness of the eigenenergy of the bound mode.

While this physics is general enough to occur in a variety of contexts, including condensed matter systems, we have considered broadband nonlinear optics as a natural system within which to discuss the implications of our results; specifically backwards sum-frequency generation.
In this context, the coherent SUSY trap becomes a potential in the spatial or temporal domain, confining optical waves to a localized region in space or time.
Such temporal trapping physics can be leveraged to enable functions that are otherwise challenging to implement.
For instance, when performing parametric amplification in the same waveguide as the SUSY trap, we may realize strong squeezing in a way that the supermode structures do not ``morph'';
this temporal confinement is in stark contrast to the case without trap, where this morphing behavior poses significant engineering challenges, e.g.,\ within quantum-optical computation.
In a broader context, besides gain, the eigenmode subspace could enable all-optical switchable cavities, delays, and filters.
Although designing, optimizing, fabricating and validating a novel device that realizes coherent trapping using backward SFG would be a challenge,
it does not require any technology or process that has not been previously demonstrated,
and we discussed an approach towards an initial, experimental proof of concept.

Furthermore, we note that the mode coupling need not be mediated only by an optical pump field: it is conceivable that a fast electro-optic drive could achieve the same features and afford a greater control of the fields.
Tunable electro-optic Bragg reflectors have recently been demonstrated \cite{SiddharthKippenberg, SongLoncar, PrencipeGallo, PohlGrange, XueQiang}:
microwave drives and transmission lines -- as used in traveling-wave electro-optic modulators \cite{ZhangLoncar} -- may be a future route to complete electro-optic control of light pulses.
Slow-wave transmission lines \cite{WangLi} could be used to minimize microwave group velocities and maximize $v_g$.
Tunable poling via $\chi^{(3)}$ could also in principle be used to achieve the same effect, via the combination of electric-field- and optically induced nonlinearities \cite{TimurdoganWatts, YanagimotoMcMahon}.

Overall, our work establishes an unexpected bridge between the field of nonlinear optics and SUSY QM, revealing a potential method to coherently trap light waves in a distinct way from
previous approaches,
with implications to optical communication and all-optical information processing, in both the realms of classical and quantum information.
The ability to slow down the propagation of light, to the point of coherently stopping and storing broad-bandwidth optical pulses, has been a research question attracting great interest, especially two decades ago \cite{YanikFan, YanikFan2, YanikSuhFan, YanikFan3, ReedSoljacicJoannopoulos}.
Nonlinear optics was, however, then seen as an insufficiently flexible and developed platform;
predominantly, temporally-modulated photonic crystals and atomic systems \cite{LiuVestergaard, PhillipsLukin} were considered more promising candidates.
Given the recent advances in integrated photonics with nonlinear- and electro-optic materials such as lithium niobate, however, these are now ostensibly prime candidates.
The experimental realization of this vision,
to stop, manipulate, and propagate light at will, via nothing more than the waveguide itself and a temporal drive -- eschewing the need for sophisticated circuitry and components in integrated photonics
-- may be finally within reach.

\vspace{\baselineskip}
{\centering \noindent\rule{0.5\normaltextwidth}{0.4pt}\par}
\vspace{\baselineskip}

\phantomsection
\addcontentsline{toc}{section}{Declarations}

\bmhead{Acknowledgments}
We thank Tathagata Banerjee, Tianyu Wang, Marc Jankowski, and Peter McMahon for helpful comments regarding the manuscript.
We thank Shi-Yuan Ma for his encouragement to finish the manuscript.

\bmhead{Funding}
We thank NTT Research for their financial and technical support.
Portions of this work were supported by the National Science Foundation (award CCF-1918549).


\pagebreak

\begin{appendices}

\setcounter{figure}{0}
\setcounter{table}{0}
\renewcommand{\thefigure}{A\arabic{figure}}
\renewcommand{\theHfigure}{A\arabic{figure}}
\renewcommand{\thetable}{A\arabic{table}}

\numberwithin{equation}{section}

\startcontents[sections] 
{
\hypersetup{linkcolor=black}
\section*{Appendix Contents}
\printcontents[sections]{}{0}{}
}

\vspace{\baselineskip}
{\centering \noindent\rule{0.5\normaltextwidth}{0.4pt}\par}
\vspace{\baselineskip}


\section{Theoretical Background and Nonlinear-optical Supersymmetric Effective Potentials}\label{sec:app:theoreticalbackground}

\begin{figure}[!htbp]
\centering
 \subfigure{\label{fig:trap:A}}
 \subfigure{\label{fig:trap:B}}
 \subfigure{\label{fig:trap:harmonicmodes}}
 \includegraphics[width=1\normaltextwidth]{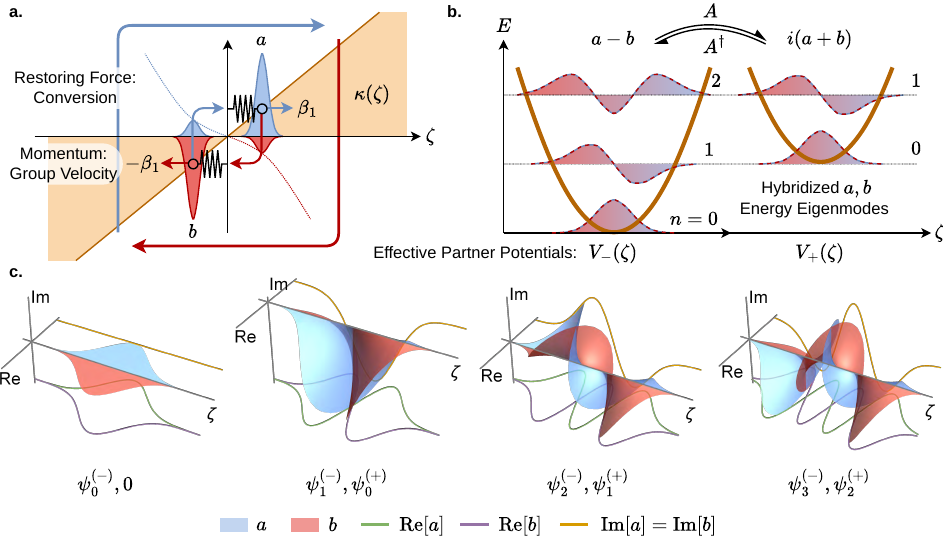}
\caption{
 \textbf{Wave trapping by mode conversion: the harmonic potential example.}
 \textbf{a.}
 Mechanics of wave trapping, based on the interplay between group velocity mismatch (GVM) and mode coupling (conversion).
 Group velocity ($\beta_1$) acts as a momentum -- the tendency of a wave mode to propagate in one direction over some coordinate $\zeta$.
 The conversion to a mode with opposite effective group velocity -- $a$ to $b$ and vice versa, mediated by a third wave or field $\kappa(\zeta)$ -- acts as a restoring force.
 These two competing effects combine to form a trap for the two modes involved.
 \textbf{b.}
 Supersymmetry picture of a harmonic wave trap.
 The trapping effect of the GVM and coupling can be formulated as a potential with trapped eigenmodes.
 Two sets of such hybridized modes emerge: $a-b$ and $i(a+b)$, with two different potentials, $V_\pm(\zeta)$, as they are intertwined by a supersymmetric relationship.
 Mathematically, these eigenmodes are interchangeable via the supercharge operators $A^{(\dagger)}$.
 \textbf{c.}
 Hybridized mode pairs of the harmonic oscillator superpotential.
 Plotted are the complex amplitudes of the first four eigenmodes for both wave modes, and their projection onto the real and imaginary axes (Re and Im).
 The eigenmodes are such that the sum and difference of the imaginary and real part, respectively, yield Hermite--Gauss functions whose order differs by one.
}
\label{fig:trap}
\end{figure}

Let us first describe our model system.
The trapping mechanism is illustrated in Fig.~\ref{fig:trap:A}.
Consider two wave modes $a,b$, propagating within some pair of coordinates $\zeta,\rho$: time and axial distance, or vice versa.
Let $a,b$ be coupled in some manner (e.g.\ by a pump), with a coupling profile $\kappa(\zeta)$.
We choose this pair of modes such that (in the frame where $\kappa(\zeta)$ is stationary) they walk off in opposite directions, due to their different effective group velocities along $\zeta$.
When a pulse is converted from one mode to the other, it will change its direction -- group velocity.
By ``bouncing'' a pulse back and forth -- by converting it each time it walks off in one direction by a given distance or delay -- we create a trap along $\zeta$ (not unlike a distributed Bragg reflector or Fabry--P\'{e}rot cavity).
We can intuitively model the walk off in either direction as momentum, and the conversion as a restoring force.
It follows that the shape of the coupling profile $\kappa$ defines an effective potential,
and we expect that a
linear profile crossing zero will like act as a linear restoring force, hence, a harmonic oscillator.

We will derive how this trapping mechanism yields an effective potential,
and, serendipitously, notice that the physics exactly follows the equations of supersymmetric quantum mechanics (SUSY QM), as illustrated in Fig.~\ref{fig:trap:B}.
This will allow us to
derive analytic expressions for the eigenmodes of arbitrary potential landscapes, and prescribe how to realize them.
Finally, we will discuss the implications and how to realize this trapping mechanism in physical, namely integrated photonic, systems.

The general model we described above is governed by the following pair of coupled wave equations:
\begin{aligned*}
\frac{da}{d\rho} &= -\kappa(\zeta) b + \beta_{1} \frac{da}{d\zeta} \\
\frac{db}{d\rho} &= \kappa(\zeta) a - \beta_{1} \frac{db}{d\zeta}
\label{eq:coupledmodes}.
\end{aligned*}
Here $a(\rho, \zeta)$ and $b(\rho, \zeta)$ represent the envelopes of the waves; we may also treat $a, b$ as a set of quantum-mechanical modes represented by field operators $\hat a(\rho, \zeta), \hat b(\rho, \zeta)$.
The cross terms proportional to the drive $\kappa$ express the coupling; and the derivative terms, with coefficient $\beta_1$, express the group velocity along $\zeta$, in the frame where $\kappa$ is stationary.
Equivalently, the mechanism is found in any system described by the following Hamiltonian $\hat H$
\begin{equation}
\hat H =
i \int \left[ \kappa(\zeta) \left( \hat{a}^\dagger \hat{b} +
\hat{a} \hat{b}^\dagger \right)
+ \beta_1\! \left( \hat{a}^\dagger \frac{d\hat{a}}{d\zeta}
- \hat{b}^\dagger \frac{d\hat{b}}{d\zeta} \right) \right]  d\zeta.
\label{eq:coupledwavehamiltonian}
\end{equation}
The only necessary conditions for the joint trapping are: that the effective group velocities have opposite signs, and that (there is a frame in which) $\kappa$ is real-valued.
For simplicity, and without loss of generality, we consider the case of equal and opposite group velocities, $\pm \beta_1$.
Appendix~\ref{sec:app:coupledwaveeq} shows how to arrive at these equations.

In order to show that this coupled-wave system supports bound modes, we would like to find the stationary states of Eq.~\ref{eq:coupledmodes}.
For this we change to a $\rho$-independent equation, using as an ansatz that the spatial evolution is simply a phase rotation $\exp(\mp i \sqrt{E} \rho)$,
and by taking even and odd superpositions (assuming real $\kappa$), we can define the following operator $A$:
\begin{aligned*}
\mp i \sqrt{E} (a - b) &=
\left(\kappa + \beta_{1} \frac{d}{d\zeta} \right) (a + b) = A^\dagger (a + b) \\
\pm i \sqrt{E} (a + b) &=
\left(\kappa - \beta_{1} \frac{d}{d\zeta}\right) (a - b)
= A (a - b) .
\label{supercharge}
\end{aligned*}
Note the resemblance to the quantum harmonic oscillator formalism that arises if we set $\kappa(\zeta) \propto \zeta$.
$\kappa$ will then act as the restoring force term, $\beta_1 d/d\zeta$ will act as the inertial term, and $A^\dagger, A$ are creation and annihilation operators.
Indeed:
\begin{aligned*}
E (a \pm b) &= 
\left\{\begin{matrix}
A A^\dagger \\
A^\dagger A
\end{matrix}\right.
(a \pm b) =
\left(-\beta_{1}^2 \frac{d^2}{d\zeta^2}
+ \kappa^2 \pm \beta_{1} \frac{d\kappa}{d\zeta}
\right) (a \pm b)
\label{traphamiltonian}
\end{aligned*}
which confirms that $a+b$ and $a-b$ are stationary states of the system.
The raising operator $A^\dagger$ maps $a-b \rightarrow \pm i(a+b)$ and, conversely, the lowering operator $A$ maps $\pm i(a+b) \rightarrow a-b$.
In the harmonic case, $\beta_{1} d\kappa/d\zeta$ is an additive constant in this effective Hamiltonian, hence $a-b$ and $i(a+b)$ are spanned by Hermite--Gauss (HG) eigenmodes, as plotted in Fig.~\ref{fig:trap:harmonicmodes}.
Individually $a,b$ are complex-valued functions whose real and imaginary parts combine to form the $n$th and $(n-1)$th HG modes.
Explicitly:
\begin{equation}
\left\{
\begin{matrix}
a - b = \text{HG}_n \\
\pm i(a + b) = \text{HG}_{n-1}
\end{matrix}
\right. \Rightarrow \left\{
\begin{array}{rrl}
\mathrm{Re}[a] =& -\mathrm{Re}[b]
&= \text{HG}_n \\
\pm \mathrm{Im}[a] =& \pm \mathrm{Im}[b]
&= \text{HG}_{n-1}
\end{array}
\right. .
\label{eq:hgeigenbasis}
\end{equation}
Note that there is a degeneracy due to the negative root of $E$, and is equivalent to exchanging $a,b \rightarrow, a^*, b^*$ \cite{taoLA}.

We have shown that wave conversion in this specific configuration can trap these wave modes as predicted,
and happened upon the equations distinctive of SUSY QM.
Indeed, in Eq.~\ref{traphamiltonian} we have rederived the SUSY QM Riccati equation \cite{CooperKhareSukhatme}:
$\kappa(\zeta)$ represents the superpotential,
and, in the general case, the (partner) potentials $V_\pm(\zeta)$ are given by:
\begin{equation}
V_\pm(\zeta) = \kappa^2 \pm \beta_1 \frac{d\kappa}{d\zeta} .
\end{equation}
Therefore, the coupling or conversion profile $\kappa$ can be engineered into a wide variety of effective potentials beyond the harmonic trap,
including ones that bind or trap a single mode.
(Note that the canonical energies here have the same units as $\rho^{-2}$.)

Let us therefore define two effective Hamiltonians, $H_\pm$, as prescribed by SUSY QM, and their real-valued eigenfunctions $\psi_{n}^{(\pm)}$,
which span the pair of hybridized modes $a-b$ and $i(a + b)$:
\begin{aligned*}
H_\pm &=
\left\{\begin{matrix}
A A^\dagger \\
A^\dagger A
\end{matrix}\right.
= -\beta_{1}^2 \frac{d^2}{d\zeta^2} + V_\pm(\zeta),
& H_\pm \psi_{n}^{(\pm)} = E_n^{(\pm)} \psi_{n}^{(\pm)}
\end{aligned*}
Note that $A$ and $A^\dagger$ are not simply creation and annihilation operators; they are the so-called supercharge operators,
as illustrated in Fig.~\ref{fig:trap:B}.
They annihilate (create) a node in the eigenfunction, as we saw in Eq.~\ref{supercharge}, and convert between eigenfunctions of $H_\pm$ and $H_\mp$ with the same energy $E$.
Explicitly \cite{CooperKhareSukhatme}:
\begin{aligned*}
E_n^{(+)} = E_{n+1}^{(-)}, \quad E_0^{(-)} = 0 \\
\psi_{n}^{(+)} = [E_{n+1}^{(-)}]^{-1/2} A \psi_{n+1}^{(-)} \\
\psi_{n+1}^{(-)} = [E_{n}^{(+)}]^{-1/2} A^\dagger \psi_{n}^{(+)} .
\label{eq:supercharge}
\end{aligned*}

Eq.~\ref{eq:supercharge}, which is valid in
unbroken SUSY (as reviewed below), has important consequences for trapping.
It tells us that in such cases, the optical trap will have a ground state in $H_-$ with zero energy.
As the energy represents the phase velocity along $\rho$,
this means that the phase-matching condition for any other process or gate performed on the ground state, e.g.\ squeezing, is not affected by the trap.
This greatly simplifies their design: such gates do not need to account for trap parameters.
Consequently, we expect gates acting on the ground state to be robust to trap depth (i.e.\ drive) noise.

Eq.~\ref{eq:supercharge}, and the zero-energy ground state property, do not apply to arbitrary superpotentials $\kappa(\zeta)$;
in those cases
it is said that SUSY is broken.
This condition is straightforward to verify:
by requiring the ground state to be annihilated (have zero energy), the resulting first-order differential equation is solved as follows
\cite{CooperKhareSukhatme}:
\begin{aligned*}
 A \psi_0 = A^\dagger \psi_0 = 0 
 \Rightarrow
 \psi_0 \propto \exp \left( \pm \beta^{-1}\! \int_0^\zeta \kappa(\zeta') d\zeta' \right) .
 \label{eq:unbrokensusy}
\end{aligned*}
Hence, the ground state $\psi_0$ exists as long as a normalizable solution exists:
in general SUSY is unbroken if $\kappa(\zeta)$ has opposite signs as $\zeta\rightarrow \pm \infty$, and broken otherwise.
In the latter case, the ground state of these traps have pump-dependent energy eigenvalues;
this case is still relevant in the context of nonlinear and quantum optics, and we consider a simple case in Appendix~\ref{sec:app:brokensusy}.

So far, for simplicity, we have solely considered the case where the two signal wavelengths have equal and opposite group velocities $\pm \beta_1$.
While the opposite sign is strictly necessary (for momentum in either direction to be possible), the equal magnitude is not.
Suppose that the group velocities for $a,b$ are associated with the propagation constants $\beta_a, \beta_b$:
\begin{aligned*}
\frac{da}{d\rho} &= -\kappa(\zeta) b + \beta_{a} \frac{da}{d\zeta}  = -\kappa(\zeta) b + \varepsilon \beta_1 \frac{da}{d\zeta} \\
\frac{db}{d\rho} &= \kappa(\zeta) a - \beta_{b} \frac{db}{d\zeta} = \kappa(\zeta) a - \frac{\beta_1}{\varepsilon} \frac{db}{d\zeta}
.
\end{aligned*}
We can recover the supercharge operators by letting
$\beta_1 = \sqrt{\beta_a \beta_b}$ and $\varepsilon = \sqrt{\beta_a / \beta_b}$,
and instead taking weighted superpositions $a \pm b \rightarrow  a / \sqrt{\varepsilon} \pm \sqrt{\varepsilon} b \propto \sqrt{|\beta_b|} a + \sqrt{|\beta_a|} b$.
In other words, the increase or reduction in the $b$ velocity is compensated via its contribution to the superposition -- a change in its ``mass'' to retain zero net velocity.
Or more simply put, a photon spends proportionately more time in the slower mode than in the faster mode.
This is analogous to the relation between group velocity and temporal profiles in two-mode squeezing \cite{UrenRaymer}.
Note that the energy ansatz in Eq.~\ref{supercharge} must be modified such that the phase velocities scale with the group velocities: $a \propto \exp(\mp i \sqrt{E} \rho / \varepsilon)$ and $b \propto \exp(\mp i \sqrt{E} \varepsilon \rho)$.

Finally, for a quantum-mechanical treatment, we can define field operators in the eigenbasis (similarly to Eq.~\ref{eq:hgeigenbasis} and Fig.~\ref{fig:trap:harmonicmodes}):
\begin{aligned*}
\hat \alpha_{\pm n} &= \frac{1}{\sqrt{2}} \int\! \left( \varphi_{\pm n} \hat a - \varphi_{\mp n} \hat b \right) d\zeta \\
\varphi_0 &= \psi_0^{(-)}, ~
\varphi_{\pm n} = \frac{1}{\sqrt{2}} \left( \psi_n^{(-)} \pm i \psi_{n-1}^{(+)} \right), \\
\text{ s.t. }&\hat H = \sum_{n, \pm} \pm\sqrt{E_n} \hat \alpha_{\pm n}^\dagger \hat \alpha_{\pm n} + \hat H_\text{free}
\end{aligned*}
where $\hat \alpha_{\pm n}$ is the $n$th energy level eigenmode bosonic field operator; as the $n > 0$ modes come in degenerate pairs, we index them by $\pm$.
If the trap is finite and the trapped modes do not form a complete basis, we must include the remainder of $\hat H$ within the Hamiltonian of the free modes, $\hat H_\text{free}$.

\subsection*{The Coupled Wave Equation}\label{sec:app:coupledwaveeq}

Here we review how to arrive at Eq.~\ref{eq:coupledmodes}.
For concreteness, we consider the nonlinear-optical, time domain coupled-wave equations,
hence we will replace $\zeta \rightarrow t$ and $\rho \rightarrow z$.
The coupled wave equations for collinear sum-frequency generation, under the slowly varying envelope approximation, and with a time-dependent pump are as follows \cite{nonlinearoptics}:
\begin{aligned*}
\frac{da}{dz} &= i \kappa(t) \mathrm{e}^{-i \Delta \beta(\omega) z} b(t) \\
\frac{db}{dz} &= i \kappa^*(t) \mathrm{e}^{i \Delta \beta(\omega) z} a(t) ,
\label{eq:sfgcoupledwave}
\end{aligned*}
where $\Delta \beta(\omega)$ is the frequency-dependent phase mismatch of the wave-mixing process.
Moving into the rotating frame ($a\rightarrow a\exp(-\Delta \beta z/2)$ and $b\rightarrow b\exp(+\Delta \beta z/2)$), and expanding the phase mismatch about the central frequency, we obtain:
\begin{aligned*}
\frac{da}{dz} &= i \kappa(t) b + i \left(\beta_{0a} + i \beta_{1a} \frac{d}{dt} - \frac12\beta_{2a} \frac{d^2}{dt^2} \ldots \right) a \\
\frac{db}{dz} &= i \kappa^*(t) a + i \left(\beta_{0b} + i \beta_{1b} \frac{d}{dt} - \frac12\beta_{2b} \frac{d^2}{dt^2} \ldots \right) b .
\end{aligned*}
The Fourier transform of the dispersion expansion replaces the angular frequency with the differential operator, $\omega \rightarrow i d/dt$.

If we consider the phase-matched and anti-symmetric linear dispersion case ($\beta_{0a} = \beta_{0b}$, $\beta_{1a} = -\beta_{1b}$, $\beta_2 = 0$),
and replace $\kappa \rightarrow i \kappa$,
we simplify to:
\begin{aligned*}
\frac{da}{dz} &= -\kappa(t) b + \beta_{1} \frac{d}{dt} a \\
\frac{db}{dz} &= \kappa^*(t) a - \beta_{1} \frac{d}{dt} b
\end{aligned*}
just as Eq.~\ref{eq:coupledmodes}, for real $\kappa(t)$.

The Hamiltonian for Eq.~\ref{eq:sfgcoupledwave} is:
\begin{equation}
H = \int d\zeta \left[ i \kappa(\zeta) \hat{a}^\dagger \hat{b} + \text{h.c.}
+ \hat{a}^\dagger \beta_a(\omega)
\hat{a}
+ \hat{b}^\dagger \beta_b(\omega)
\hat{b} \right] ,
\end{equation}
which yields Eq.~\ref{eq:coupledwavehamiltonian} after performing the same transformations and truncation to linear dispersion.

\section{The OPA Interaction Hamiltonian}\label{sec:app:opahamiltonian}

The Hamiltonian for two-mode parametric amplification with a drive $\gamma$ is:
\begin{equation}
\hat H_\text{OPA} = i \int \left( \gamma(\zeta) \hat a^\dagger \hat b^\dagger + \gamma^*(\zeta) \hat a \hat b \right) d\zeta
\end{equation}
and would contribute the terms $i \gamma(\zeta) b^*$ and $i \gamma(\zeta) a^*$ to the equations of motion given in Eq.~\ref{eq:coupledmodes}.

Solving for the local field operators with respect to the trapped modes, we have:
\begin{aligned*}
\hat a(\zeta) = & \frac{1}{\sqrt{2}} \sum_{n, \pm} \varphi_{\pm n}(\zeta) \hat \alpha_{\pm n} + \sum_m \psi^{(a)}_{m,\text{free}}(\zeta) \hat a_{m,\text{free}} \\
\hat b(\zeta) = - & \frac{1}{\sqrt{2}} \sum_{n, \pm} \varphi_{\mp n}(\zeta) \hat \alpha_{\pm n} + \sum_m \psi^{(b)}_{m,\text{free}}(\zeta) \hat b_{m,\text{free}},
\end{aligned*}
where we may include terms for the higher-order free modes for the case of finite potentials.

We can thus write the $\hat H_\text{OPA}$ in the interaction picture (for simplicity we omit the higher-order free modes):
\begin{aligned*}
\mathrm{e}^{i \hat H \rho} \hat H_\text{OPA} \mathrm{e}^{-i \hat H \rho}
&= -\frac12 \mathrm{e}^{i\hat H \rho} \left[ \int \gamma^*(\zeta)
\left( \sum_{n, \pm} \varphi_{\pm n}(\zeta) \hat \alpha_{\pm n} \right)
\left( \sum_{n', \pm} \varphi_{\mp n'}(\zeta) \hat \alpha_{\pm n'} \right)
d \zeta + h.c. \right] \mathrm{e}^{-i\hat H \rho} \\
&= -\frac12 \mathrm{e}^{i \hat H \rho} \left( \sum_{\pm n, \pm n'} \int \gamma^* \varphi_{\pm n} \varphi_{\mp n'} d\zeta\ \hat \alpha_{\pm n} \hat \alpha_{\pm n'} + h.c. \right) \mathrm{e}^{-i \hat H \rho}
\end{aligned*}

In the case of constant coupling, $\gamma(z) = \gamma_0$, and because $\varphi_{\pm n}^* = \varphi_{\mp n}$, the overlap integral naturally excludes combinations of different eigenfunctions, and allows us to simplify:
\begin{equation}
-\frac{\gamma_0^*}{2} \mathrm{e}^{i\hat H \rho} \left( \hat \alpha_0^2 + \sum_{n, \pm} \hat \alpha_{\pm n} \hat \alpha_{\pm n} \right) \mathrm{e}^{-i\hat H \rho} + h.c.
\end{equation}
hence, the ground state and all higher-order eigenfunction are independently squeezed, without entanglement to each other or to free modes.
In other words, they correspond to squeezing supermodes.
As per the exponentials, they induce rotations with frequency $\pm\sqrt{E_n}$, hence only the ground mode is phase matched.
\begin{equation}
-\frac{\gamma_0^*}{2} \left( \hat \alpha_0^2 + \sum_{n} \hat \alpha_{\pm n}^2 \mathrm{e}^{\mp 2i \sqrt{E} \rho} \right) + h.c.
\end{equation}

A finite-bandwidth $\gamma(\zeta)$ gives rise to (non-phase-matched) eigenmode cross-terms -- hence the supermodes and eigenmodes may not correspond.

Curiously, if we instead consider single-mode squeezing in the physical modes (degenerate OPA), i.e. $\hat H_\text{OPA} \sim \hat a^2$ and/or $\hat b^2 + h.c.$, we obtain two-mode squeezing in the eigenmodes:
\begin{equation}
\gamma_0^* \left( \frac{\hat \alpha_0^2}{2} + \sum_{n} \hat \alpha_{\pm n} \hat \alpha_{\mp n} \right) + h.c.
\end{equation}
In this case, since the $\pm$ eigenmodes have opposite phase velocity, the squeezing terms for all the pairs of modes are phase matched.
This is also true in the previous OPA scenario, in the general case of arbitrary $\gamma(z)$: all modes may couple to all modes, however, only these $\pm$ pairs are phase matched.
In this case, free modes may also play a role, if applicable.

We note that in the degenerate OPA case, the phase matching over all modes is to be expected since the group velocity dispersion was assumed to be negligible, and therefore the phase-matching bandwidth will be arbitrarily broad \cite{WasilewskiRadzewicz}.

\subsection*{Hybridization Under Broken Supersymmetry and Parametric Amplification of Free Modes}\label{sec:app:brokensusy}

In this section we describe the mode hybridization that occurs under a constant coupling, as a representative example of when supersymmetry is broken (Eq.~\ref{eq:unbrokensusy} is not satisfied).
We then note how a simultaneous (perturbative) parametric amplification process is influenced  by the presence of coupling.
This is relevant when considering the free modes not bound by the effective potentials, far from the node of the pump.
For simplicity, we will continue use the language of nonlinear optics.

Sum-frequency generation (SFG) is described by the coupled wave equations above (Eq.s~\ref{eq:sfgcoupledwave}).
In the rotating frame, the equations of motion can be rewritten as:
\begin{equation}
\frac{d}{dz} \left(\begin{matrix}
a\\b
\end{matrix} \right) =
i \left(\begin{matrix}
\tfrac12 \Delta \beta & \kappa \\
\kappa^* & -\tfrac12 \Delta \beta
\end{matrix} \right)
\left(\begin{matrix}
a\\b
\end{matrix} \right) .
\end{equation}
Generally, SFG couples -- hence hybridizes -- the modes $a,b$.
Indeed the eigenvalues of this Hamiltonian are
\begin{equation}
E^{(\pm)} = \pm \frac12 \sqrt{\Delta \beta(\omega)^2 + 4\kappa^2} .
\end{equation}
When phase-matched, i.e.\ $\Delta \beta(\omega) = 0$, the eigenstates of evolution are the sum and difference of the two modes:
$(a \pm b) / \sqrt{2}$.
However, with a constant $\kappa$ profile, neither eigenvalue $E^{(\pm)}$ vanishes to give zero energy, which can be interpreted as broken supersymmetry.

The dispersion relation for these hybridized modes with constant coupling is of interest.
These are relevant when considering free modes, or modes far from the potential well, e.g. in the tanh superpotential, which converges to a constant value: $\kappa(\rightarrow \pm \infty) \rightarrow \text{const}$.
As the dispersion is nonlinear, these joint modes have the tendency to broaden, due only to the group velocity mismatch;
the pulses in $a,b$ walk off, while undergoing Rabi cycles.

Similarly to the SFG process, an optical parametric amplifier (OPA), which acts on two modes $a$ and $b$, is described by the equations:
\begin{aligned*}
\frac{da}{dz} &= i \gamma b^* \mathrm{e}^{i \Delta \beta_\text{OPA}(\omega) z} \\
\frac{db}{dz} &= i \gamma a^* \mathrm{e}^{i \Delta \beta_\text{OPA}(\omega) z}
\end{aligned*}
for some phase mismatch $\Delta \beta_\text{OPA}(\omega)$ and coupling $\gamma$.
Notably, these equations share the eigenstates with the SFG process in the phase-matched case, $\Delta \beta_\text{OPA} = 0$.
This implies that SFG and OPA can be combined, in order to squeeze the dressed states $(a \pm b) / \sqrt{2}$.
This requires the SFG Hamiltonian to dominate, hence $\kappa \gg \gamma$.

Let us now consider the broadband, frequency-multimode case.
We are interested in the case of a broadband OPA pump, but a monochromatic SFG pump.
For simplicity, as before, we consider the linear dispersion case, where the wavenumbers of the two modes scale as:
$\beta_{a/b}(\omega_0 + \delta \omega) = \beta(\omega_0) \pm \beta_{1} \delta \omega$.
Therefore, we write the SFG Hamiltonian for $\Delta \beta=0$ and then transform it into the dressed basis:
\begin{aligned*}
H_\text{SFG}(\omega_i, \omega_i') &=
\left(\begin{matrix}
\beta_{1} \omega_a ~\delta(\omega_a-\omega_a') & \kappa ~\delta(\omega_b-\omega_a') \\
\kappa^* ~\delta(\omega_b-\omega_b') & -\beta_{1} \omega_b ~\delta(\omega_b-\omega_b')
\end{matrix}\right)
\\
\rightarrow
H_\text{SFG}(\omega, \omega') &=
\left(\begin{matrix}
\sqrt{\beta_1^2 \omega^2 + \kappa^2} ~\delta(\omega-\omega') & \mathbb{0} \\
\mathbb{0} & -\sqrt{\beta_1^2 \omega^2 + \kappa^2} ~\delta(\omega-\omega')
\end{matrix}\right)
\end{aligned*}
(we include the Dirac delta function $\delta (\omega)$ to indicate that the SFG pump is monochromatic).
Call the hybridized eigenstates:
\begin{equation}
c = \frac{a + b}{\sqrt{2}}, ~
d = \frac{a - b}{\sqrt{2}}
\end{equation}
and we can substitute these into the OPA Hamiltonian:
\begin{aligned*}
H_\text{OPA} = \gamma(\omega_a + \omega_b) a^\dagger(\omega_a) b^\dagger(\omega_b) & \mathrm{e}^{i \Delta \beta_\text{OPA}(\omega_a, \omega_b) z} \\
= \tfrac12 \gamma(\omega_a + \omega_b)
\mathrm{e}^{i \Delta \beta_\text{OPA} z}
&\left( c^\dagger(\omega) \exp\left(-i z \sqrt{\beta_1^2 \omega^2 + \kappa^2}\right) + d^\dagger(\omega) \exp\left(i z \sqrt{\beta_1^2 \omega^2 + \kappa^2}\right) \right) \\
&\left( c^\dagger(\omega') \exp\left(-i z \sqrt{\beta_1^2 \omega'^2 + \kappa^2}\right) + d^\dagger(\omega') \exp\left(i z \sqrt{\beta_1^2 \omega'^2 + \kappa^2}\right) \right) .
\end{aligned*}
Consider, for example, the $c^\dagger(\omega) c^\dagger(\omega')$ term.
The term rotates as
\begin{equation}
\beta_\text{tot}(\omega, \omega') = \Delta \beta_\text{OPA} - \sqrt{\beta^2 \omega^2 + \kappa^2} - \sqrt{\beta^2 \omega'^2 + \kappa^2} ,
\end{equation}
which depends on the pump strength $\kappa$.
The phase matching condition at the central frequency may be satisfied when
$\Delta \beta_\text{OPA} \approx 2 |\kappa|$.
Notice also that there is a level splitting: the $d^\dagger(\omega) d^\dagger(\omega')$ term in the Hamiltonian
would insteada be phase matched at $\Delta \beta_\text{OPA} \approx -2|\kappa|$.
In these cases, when $\kappa / \beta_1 \gg \omega, \omega'$, the phase rotation term is:
\begin{aligned*}
\beta_\text{tot}(\omega, \omega') &=
- \frac{\beta_1^2}{2\kappa} \left( \omega^2 + \omega'^2 \right) .
\end{aligned*}
Interestingly, when the quadratic approximation holds, $\beta_\text{tot}$ has circular symmetry, which implies separability.


\end{appendices}

\phantomsection
\addcontentsline{toc}{section}{References}

\putbib[%
 refs%
]
\restoregeometry
\end{bibunit}

\end{document}